\documentclass[aps,reprint,prl,showpacs,amsmath,amssymb, nobibnotes]{revtex4-1}
\usepackage{hyperref}
\usepackage{epsfig}
\usepackage[utf8]{inputenc}
\usepackage{bbold,amsfonts}
\usepackage{graphicx}
\usepackage{amssymb,amsmath}
\usepackage{fancybox,framed}
\usepackage{dsfont}
\usepackage{mathtools}
\usepackage{braket}
\usepackage{slashed}
\usepackage{rotating}

\newcommand{\beq}{\begin{equation}}
\newcommand{\eeq}{\end{equation}}

\renewcommand{\a}{\alpha}

\renewcommand{\d}{\delta}
\newcommand{\pa}{\partial}

\newcommand{\e}{\epsilon}

\renewcommand{\l}{\lambda}

\newcommand{\p}{\pi}

\newcommand{\s}{\sigma}
\renewcommand{\S}{\Sigma}
\renewcommand{\t}{\tau}

\begin{document}

\title{Marginal deformations and defect anomalies}%

\author{Lorenzo Bianchi}%
\affiliation{Center for Research in String Theory - School of Physics and Astronomy Queen Mary University of London, Mile End Road, London E1 4NS, UK}

\begin{abstract}
We deform a defect conformal field theory by an exactly marginal bulk operator and we consider the dependence on the marginal coupling of flat and spherical defect expectation values. For odd-dimensional defects, we find a different qualitative behavior for the flat and spherical case, generalizing to arbitrary dimensions the line-circle anomaly of superconformal Wilson loops. In the even-dimensional case, on the other hand, we find a logarithmic divergence which can be related to a $a$-type anomaly coefficient. This coefficient, for defect theories, is not invariant on the conformal manifold and its dependence on the bulk coupling is controlled to all orders by the one-point function of the associated exactly marginal operator. In particular, our results imply a non-trivial dependence on the bulk coupling for the recently proposed defect $C$-function. We finally apply our general result to a few specific examples, including superconformal Wilson loops and R\'enyi entropy.
\end{abstract}


\maketitle

\section{Introduction and results}
Extended probes play a distinguished role in a wide range of physical phenomena. Wilson and 't Hooft lines, boundaries, interfaces and twist operators provide physically interesting examples of a broad class of observables denoted as defects. In the hope of identifying universal properties, it is convenient to restrict our attention to non-local operators preserving conformal symmetry along their profile. The study of conformal defects started a long time ago in two dimensions \cite{Cardy:1984bb,McAvity:1993ue}, but only recently the constraints of conformal symmetry have been systematically imposed in higher dimensions \cite{Billo:2016cpy}.

Among all the examples of conformal defects, superconformal Wilson lines provide an extremely useful laboratory, both because we have lot of data at our disposal and because of the variety of techniques that can be used to access their non-perturbative regime. The most famous example is certainly the Maldacena Wilson loop in $\mathcal{N}=4$ Super Yang Mills theory \cite{Maldacena:1998im}. In that case, the expectation value for the circular Wilson loop is known to all orders in the coupling \cite{Erickson:2000af,Drukker:2000rr,Pestun:2007rz} and it is different from the straight line expectation value, despite the two configurations are mapped to each other by a conformal transformation. This fact can be attributed to a conformal anomaly in the transformation relating the straight line and the circle \cite{Drukker:2000rr}. Similar phenomena have been observed in all those cases where the expectation value of the circular Wilson loop could be computed exactly, such as $\mathcal{N}=2$ theories in four dimensions \cite{Pestun:2007rz,Passerini:2011fe} or $\mathcal{N}\geq2$ theories in three dimensions \cite{Kapustin:2009kz,Griguolo:2013sma,Bianchi:2013rma,Mauri:2018fsf}. It is therefore a natural question whether such an anomaly is a more general feature of conformal defects, i.e. whether it is always true that the flat defect expectation value is different from the spherical one. More generally, one may wonder whether it makes sense to compute the defect expectation value in the flat case where the only available scales are the IR and UV cut-off (we will say more on this point in the following). In this letter, we argue that flat and spherical defects exhibit indeed different qualitative features and that, even assuming one could make sense of the flat defect expectation value, the spherical one, in general, is different and it is a non-trivial function of the marginal coupling. Let us stress, however, that the plane-sphere anomaly is not related to any geometric invariant and it is qualitatively different from the more familiar case of the defect Weyl anomaly.

The Weyl anomaly is an important feature of homogeneous conformal field theories (CFT) in even dimensions. A way to describe such anomaly is through the expectation value of the trace of the stress tensor $T^{\mu}{}_{\mu}$. When the theory is embedded in an arbitrary curved manifold $T^{\mu}{}_{\mu}$ acquires a non-vanishing expectation value proportional to a linear combination of Weyl invariants. The coefficients of this linear combination are called anomaly coefficients and their number grows with the spacetime dimension. It is a well-known fact that for $2$ and $4$ dimensions all the Weyl anomaly coefficients can be related to pieces of conformal data, in particular to the two- and three-point functions of the stress tensor \cite{Osborn:1993cr}. 

In the presence of defects, the number of geometric invariants grows significantly, given the presence of additional ingredients like the extrinsic curvature of the defect profile. The simplest possible case is that of a two-dimensional surface, for which the relevant invariants have been classified \cite{Graham:1999pm}. In that case there is one $a$-type and two $b$-type anomaly coefficients. Interestingly, the two $b$-type coefficient could be mapped to the one-point function of the stress tensor operator and the two-point function of the displacement operator \cite{Bianchi:2015liz}. On the contrary, the $a$-type coefficient, which is particularly interesting for its expected monotonicity property under RG flow \cite{Jensen:2015swa,Kobayashi:2018lil}, has not been related to any piece of defect conformal data yet.

In this letter, we study a particular class of conformal field theories characterized by the presence of a scalar operator of protected dimension $d$, i.e. an exactly marginal operator. In that case the action can be deformed by
  \begin{equation}\label{marginal}
  S\to S+\l\int d^d x\  \mathcal{O}(x)
 \end{equation}
 where we assumed the CFT lives in flat space and for simplicity we restrict to a single marginal direction. The argument easily generalizes to the case of several marginal operators.
Examples of such theories are very common in the presence of supersymmetry in four dimensions and less common in three dimensions (see \cite{Cordova:2016xhm} for a full classification), but the existence of non-supersymmetric conformal manifolds in $d>2$ is an interesting open question. 

We consider the dependence of the defect expectation value on the marginal coupling $\l$ associated to the exactly marginal operator. For the case of a flat defect, the result is expected to be both IR and UV divergent and the two cut-offs are the only available scales. It is therefore not clear whether one could identify a part of the result which is independent of the regularization scheme in order to make sense of it for general defects. For the Wilson line, there is a well-defined renormalization procedure that goes back a long time \cite{Dorn:1980hs,Aoyama:1981ev} (a more recent discussion can be found in \cite{Griguolo:2012iq}). Our result shows that, even if one could identify a universal part for the flat defect, it will always be independent of the marginal coupling. 

For a spherical defect the situation is more interesting. The dependence on the marginal coupling is controlled by the one-point function of the exactly marginal operator 
\begin{equation}\label{flat1pt}
 \braket{\mathcal{O}(x)}_{\text{plane}}=\frac{C_O}{|x_{\perp}|^{d}}
\end{equation}
and, as expected, it behaves differently for even and odd dimensional defects. In particular, for a spherical conformal defect $\S$ of dimension $p$ and codimension $q$ such that $p+q=d$, we find that 
\begin{align}
 \pa_{\l} \log \braket{\S}= \left\{ \begin{array}{ll} C_O \frac{(-1)^{\frac{p}{2}+1} 4 \pi^{\frac{p+q}{2}}}{\Gamma(\frac{p}{2}+1)\Gamma(\frac{q}{2})} \log \e & \text{even } p\\ 
                                      C_O \frac{(-1)^{\frac{p+1}{2}} 2 \pi^{\frac{p+q}{2}+1}}{\Gamma(\frac{p}{2}+1)\Gamma(\frac{q}{2})} & \text{odd } p
                                     \end{array}\right.
\end{align}
where $\e$ is a UV cut-off and the sphere radius has been set to one.
This is in agreement with general expectations, since we know that for even $p$ the universal part of the defect expectation value is given by the coefficient of the logarithmic divergence. In particular, this coefficient can be expressed as a linear combination of Weyl invariants and the expectation value of the spherical defect is related to the $a$-type anomaly. More generally, the spherical defect expectation value has been identified as the best candidate for a monotonically decreasing function under defect RG flow both in even and odd dimensions, i.e. a $C$-function \cite{Kobayashi:2018lil} (see also \cite{Nozaki:2012qd,Gaiotto:2014gha,Estes:2014hka,Yamaguchi:2002pa} for previous theorems and conjectures encoded by the proposal in \cite{Kobayashi:2018lil}). Here we find that this proposed defect $C$-function depends non-trivially on the bulk marginal couplings. This is no contradiction with usual results since under a defect RG-flow a candidate defect $C$-function should not depend on defect marginal couplings, but could depend in general on bulk marginal couplings. 

\textbf{Note added:} While this paper was in the final stage of its preparation, the preprint \cite{Herzog:2019rke} appeared on the arXiv, which overlaps with this work on several aspects. The authors of \cite{Herzog:2019rke} considered the partition function of a spherical defect in a spherical background and derived the Wess-Zumino consistency conditions for the conformal anomaly. Here we consider the expectation value of a spherical defect in flat space and we include some explicit examples.

\section{Conformal defects and marginal deformations}
  We start by considering the derivative of the logarithm of the defect expectation value with respect to the exactly marginal coupling. It is immediate to see that this derivative is related to the one-point function of the exactly marginal operator $\braket{\mathcal{O}(x)}_{\S}$ \footnote{One may be worried of a possible explicit dependence on $\l$ in the definition of the defect. If this was the case, the first-order derivative with respect to the marginal coupling would give the one-point function of a defect operator, which vanishes by conformal invariance} 
\begin{equation}\label{derivative}
 \pa_{\l}\log\braket{\S}=\int d^d x \braket{\mathcal{O}(x)}_{\S}
\end{equation}
where the defect one-point function is normalized by the defect vacuum
\begin{equation}
 \braket{\mathcal{O}(x)}_{\S}=\frac{\braket{\mathcal{O}(x)\S}}{\braket{\S}}
\end{equation}
Importantly, the kinematics of this one-point function is completely fixed by conformal invariance and the dynamical content is encoded in a single constant $C_O$, which would depend, however, on all the free parameters in the bulk and defect theory and, in particular, on the marginal parameter $\l$. The explicit expression for the one-point function in flat space is trivial and was given in \eqref{flat1pt}. 
Also the extension to the spherical case is not particularly difficult. Actually, it is particularly simple using the embedding formalism developed in \cite{Billo:2016cpy}, where one just needs to pick two different sections of the projective cone. The result is that, for the case of a spherical defect of unit radius we can split the Euclidean coordinates into $p+1$ ``parallel'' and $q-1$ ``orthogonal'' coordinates
\begin{equation}
 x^{\mu}=(x_{\parallel}^{\hat{i}},x_{\perp}^{\hat a})
\end{equation}
where we used quotation marks to highlight that the parallel coordinates, labeled by an index $\hat{i}$, are the $p+1$ directions in which the sphere is embedded. Consistently the index $\hat{a}$ runs over the $q-1$ orthogonal directions. Specifically, the sphere is defined by
\begin{equation}\label{coordsphere}
 x_{\parallel}^{\hat{i}} x_{\parallel \hat{i}}=1
\end{equation}
In this coordinate system, the one-point function in presence of a spherical defect reads
\begin{equation}\label{sphere1pt}
 \braket{\mathcal{O}(x)}_{\text{sphere}}=\frac{C_O}{\left(x_{\perp}^2+\frac{(1-x_{\perp}^2-x_{\parallel}^2)^2}{4}\right)^{d/2}}
\end{equation}
Despite quite unusual, this system of coordinates is particularly useful to perform the integration in \eqref{derivative}.

\subsection{A trivial case: the flat defect}
The first question we would like to address is what happens if one performs the integral \eqref{derivative} in the case of a flat defect. The answer is quite trivial. The flat defect case is IR and UV divergent and its expectation value, in general, is bound to suffer from these divergences. If one tries to regulate the integral \eqref{derivative} with a IR cutoff $L$ and a UV cutoff $\e$, the result is
\begin{align}\label{flatint}
 \pa_{\l}\log\braket{\S}&=C_O S_{q-1} S_{p-1}\!\! \int_\e^\infty\!\!\!\! dr_{\perp} \frac{1}{r_{\perp}^{d-q+1}}  \int_0^L \!\!\!dr_{\parallel} r_{\parallel}^{p-1} \nonumber\\
 &=C_O S_{q-1} S_{p-1} \left(\frac{L}{\e}\right)^p 
\end{align}
 where $S_{n}=2\frac{\p^{\frac{n+1}{2}}}{\Gamma(\frac{n+1}{2})}$ is the volume of the $n$-sphere. The result \eqref{flatint} for integer values of $p$ is a power-like divergence and therefore it is an artifact or the regularization procedure. In particular, one can consider performing the integral in dimensional regularization by allowing $p$ to take non-integer values. In that case, the integrals appearing in \eqref{flatint} are the typical examples of scaleless integrals and therefore must be set to zero in dimensional regularization. The correct way to interpret this result is to affirm that there is no universal part of the flat defect expectation value which depends on the marginal coupling. In other words, even if some symmetry protects the flat expectation value from the aforementioned divergences or if one could identify a universal part after renormalization, the result would not depend on $\l$. In the literature, similar issues are discussed in the context of supersymmetric Wilson lines \cite{Griguolo:2012iq}.
 
\subsection{The spherical defect}
We will see that the story is quite different for the case of a spherical defect. Using the coordinates introduced in \eqref{coordsphere} and trivially performing the angular integrations we obtain
\begin{equation}\label{sphereintegral}
 \pa_{\l}\log\braket{\S}=C_O S_{q-2} S_{p} \int  \frac{dr_{\perp}dr_{\parallel} r_{\perp}^{q-2}r_{\parallel}^{p}}{\left(r_{\perp}^2+\frac{(1-r_{\perp}^2-r_{\parallel}^2)^2}{4}\right)^{d/2}}
\end{equation}
with $r_{\parallel}=|x_{\parallel}|$ and $r_{\perp}=|x_{\perp}|$. The location of the divergence is geometrically very clear, since the defect is positioned at $r_{\perp}=0$ and $r_{\parallel}=1$. As a double check of our result, we perform the integrals in two different ways.

\subsubsection{Dimensional regularization}
In the first example, we will use a defect version of dimensional regularization such that $d=p+q$ is fixed, but $p$ and $q$ are kept generic. That means we are not changing the total dimension of the space, but we are changing defect dimension and codimension. Introducing polar coordinates $(r_{\parallel}=r \sin \theta,r_{\perp}=r \cos \theta)$ we get
\begin{align}
 &\pa_{\l}\log\braket{\S}=  \\&C_O S_{q-2} S_{p} \int_0^{\infty} dr \int_0^{\frac{\pi}{2}} d\theta \frac{(\cos\theta)^{q-2}(\sin\theta)^{p}}{r\left(\cos\theta^2+\frac{(\frac{1}{r}-r)^2}{4}\right)^{p+q/2}} \nonumber
\end{align}
and, after performing the integrals
\begin{align}
\pa_{\l}\log\braket{\S}=C_O \frac{2 \pi^{\frac{d}{2}}\Gamma(-\frac{p}{2})}{\Gamma(\frac{q}{2})}
\end{align}
Clearly this expression has single poles for even values of $p$. This is expected since for even dimensional defects the universal part of the free energy is proportional to a logarithmic UV singularity. To extract the coefficient of the logarithm we can simply take the residue 
\begin{align}\label{evenpresult}
 \pa_{\l}\log\braket{\S}_{\text{univ}}^{\text{even }p}=C_O\frac{(-1)^{\frac{p}{2}+1} 4 \pi^{\frac{d}{2}}}{\Gamma(\frac{p}{2}+1) \Gamma(\frac{q}{2})}
\end{align}
On the other hand, for odd $p$ the result is finite and it is simply
\begin{align}\label{oddpresult}
 \pa_{\l}\log\braket{\S}^{\text{odd }p}=C_O \frac{2 \pi^{\frac{d}{2}}\Gamma(-\frac{p}{2})}{\Gamma(\frac{q}{2})}
\end{align}
We will see that these results are perfectly reproduced by a cut-off analysis, where the physical interpretation of the divergence is more transparent.

\subsubsection{Cut-off regularization}
 We compute the integral \eqref{sphereintegral} for integer values of $p$ and $q$ regularizing by a UV cutoff $\e$ around $r_{\parallel}=1$, where the defect is located. We then isolate the universal term in the expansion (i.e. the term that is not affected by a rescaling of the cutoff). For even $p$ it appears as the coefficient of $\log \e$, while for odd $p$ it is a finite part. 

Let us start by even $p$ and $q>1$. One can perform the integral for several integer values of $p$ and $q$ and then extract the coefficient of the logarithmic singularity. Doing so, one finds they fit in the pattern
\begin{equation}
 \pa_{\l}\log\braket{\S}^{\text{even }p}_{\text{univ}}=C_O \frac{(-1)^{\frac{p}{2}+1} 4 \pi^{\frac{p+q}{2}}}{\Gamma(\frac{p}{2}+1)\Gamma(\frac{q}{2})} \log \e
\end{equation}
in perfect agreement with \eqref{evenpresult} (in this case however we left the logarithm explicit since it is related to the actual value of the cutoff).
 
For odd $p$ and $q>1$ a similar analysis gives
\begin{equation}
 \pa_{\l}\log\braket{\S}^{\text{odd }p}_{\text{sphere}}=C_O \frac{(-1)^{\frac{p+1}{2}} 2 \pi^{\frac{p+q}{2}+1}}{\Gamma(\frac{p}{2}+1)\Gamma(\frac{q}{2})}
\end{equation}
which agrees with \eqref{oddpresult} for odd $p$.

When $q=1$ we do not have orthogonal coordinates in the spherical case (let us stress again that we denote as parallel all the directions where the sphere is embedded) and one may be worried that the previous results do not apply to this specific case. Actually, in this case the integral is simply 
\begin{equation}
 \pa_{\l}\log\braket{\S}^{q=1}_{\text{sphere}}=C_O S_{p} 2^d \int dr_{\parallel} \frac{r_{\parallel}^p}{\left|1-r_{\parallel}^2\right|^{p+1}}
\end{equation}
and we find that its universal part agrees perfectly with \eqref{oddpresult} and \eqref{evenpresult} evaluated at $q=1$. Notice however that this is true only if the integral extends from zero to infinity, so in the case of a codimension one defect (not a boundary, and not an interface with two different theories on the two sides).  

\section{Examples}
In this section, we apply our general result to a few interesting examples. First, we specify our formula to line defects and we extract predictions for the Lagrangian expectation value of three- and four-dimensional theories where the Wilson loop is known exactly. For the case of $\mathcal{N}=4$ SYM we are particularly lucky because the Lagrangian expectation value can be computed independently thus confirming the validity of our formula. Then we consider the case of a two-dimensional defect, where the set of independent Weyl invariants is well known. This allows us to write down an equation for the $a$-type anomaly coefficient and to show that it is related to the scalar one-point function $C_O$. Finally, we consider the general case of codimension 2 that is relevant for the twist operator, i.e. the operator whose expectation value computes the R\'enyi entropy. For the four-dimensional case, using a specific feature of the twist operator, we can write down a general formula relating the one-point function of the stress-tensor to $C_O$.

\subsection{Line defects}
If we specify our formula \eqref{oddpresult} to the case $p=1$ we find
\begin{equation}
 \pa_{\l} \braket{\mathcal{W}}=-2\pi S_{q-1} C_O
\end{equation}
where we used $\mathcal{W}$ to denote the line defect. We will consider the four-dimensional case, where we can perform explicit checks for the validity of this formula. Let us point out, however, that marginal deformations exist also for three-dimensional $\mathcal{N}=2$ SCFT, where an exact expression for the Wilson loop is available \cite{Kapustin:2009kz}. We leave further analysis in this direction for future work.

\subsubsection{Wilson loops in four-dimensional SCFT}

Exactly marginal deformations exist for four-dimensional theories with $\mathcal{N}=1$, $\mathcal{N}=2$ and $\mathcal{N}=4$ supersymmetry. For all these cases we have
\begin{align}
   C_O&=-\frac{1}{8\pi^2} \pa_{\l} \braket{\mathcal{W}} 
\end{align}
Focusing on the case of Super Yang-Mills theories, where the marginal coupling is precisely the Yang-Mills coupling, it is convenient to perform a change of variables. In \eqref{derivative} we assumed that the coupling $\l$ multiplies the marginal operator, while in the ordinary Yang-Mills lagrangians the inverse of the 't Hooft coupling appears in front of the action. Let us therefore introduce the coupling $\hat{\l}=\frac{1}{\l}$ and reabsorb a factor of $\hat {\l}$ in the definition of $C_O$ such that $\hat{\l}\hat{C}_O= C_O$. This gives
\begin{align}
   \hat{C}_O&=\frac{1}{8\pi^2} \hat{\l}\pa_{\hat{\l}} \braket{\mathcal{W}} \label{4dline}
\end{align}
which can be used to extract predictions for the one-point function of the Lagrangian for those theories where the exact expression of the Wilson loop is known. This is perhaps not very surprising, but for $\mathcal{N}=4$ SYM it constitutes a strong consistency check since the one-point function of the Lagrangian can be computed independently. In the following we will review the argument which already appeared in \cite{Fiol:2015mrp}.

The expectation value of the circular Wilson loop in $\mathcal{N}=4$ SYM is given by~\cite{Erickson:2000af,Drukker:2000rr,Pestun:2007rz}
\begin{equation}
 \braket{\mathcal{W}}=\frac{1}{N}L^1_{N-1}\left(-\frac{\hat{\l}}{4N}\right) e^{\frac{\hat{\l}}{8N}}
\end{equation}
where $L$ is the modified Laguerre polynomial. To obtain an independent prediction on the Lagrangian one-point function (without taking an explicit derivative with respect to $\hat{\l}$) we can do the following. First of all, we consider the Bremsstrahlung function, defined as the coefficient of the second order term in the small angle expansion of the cusp anomalous dimension \cite{Correa2012}
\begin{align}
 \Gamma_{\text{cusp}}(\phi)&=-B(\hat{\l},N)\phi^2+\mathcal{O}(\phi^4) 
\end{align}
This function can be computed exactly combining defect techniques with supersymmetric localization \cite{Correa2012}
\begin{align}\label{exactB}
B(\hat{\l}, N)=\frac{1}{2\pi^2}{\hat{\l}} \pa_{\hat{\l}}  \braket{\mathcal{W}}
\end{align}
Then, we introduce the one-point function of the stress tensor (for a straight Wilson line in direction 4 and $n_a=\frac{x_{a}}{|x_{\perp}|}$ with $a=1,2,3$)
\begin{align}
 \braket{T_{ab}}&=-\frac{h({\hat{\l}},N)}{x_{\perp}^4} (\d_{ab}-2n_a n_b)\,, \\
 \braket{T_{a4}}&=0\,, \qquad \braket{T_{44}}_W=\frac{h({\hat{\l}},N)}{x_{\perp}^4}\,,
\end{align}
and its relation to the Bremsstrahlung function \cite{Lewkowycz:2013laa, Bianchi:2018zpb} 
\begin{equation}
  h({\hat{\l}},N)=\frac{B({\hat{\l}},N)}{3}
\end{equation}
Finally, we use the fact that in $\mathcal{N}=4$ SYM the Lagrangian and the stress tensor operator belong to the same supermultiplet. Using supersymmetric Ward identities along the lines of \cite{Gomis:2008qa,Fiol:2015spa,Bianchi:2018zpb} one can prove that 
\begin{equation}
 h({\hat{\l}},N)=\frac43 \hat{C}_O ({\hat{\l}},N)
\end{equation}
The same relation, in slightly different conventions, was found in \cite{Fiol:2012sg}. Therefore, combining all these results we get
\begin{equation}
  {\hat{\l}} \pa_{\hat{\l}} \log \braket{\mathcal{W}}=8\pi^2 \hat{C}_O(\hat{\l},N)
\end{equation}
in perfect agreement with formula \eqref{4dline}. Using the same reasoning backwards, i.e. starting from the validity of \eqref{4dline}, we obtain an alternative derivation of the exact formula for the Bremsstrahlung function \eqref{exactB} without using the localization argument of \cite{Correa2012}.

\subsection{Surface defects}
As we mentioned, for even dimensional defects, our result relates the one-point function of the exactly marginal operator to the derivative with respect to the coupling of the defect $a$-anomaly. For the case $p=2$, the general form of the universal part of the defect free energy  is given by 
\begin{equation}
 \left.\log \braket{\S}\right|_{\text{univ}}=\left(\frac{f_a}{2\pi} I_a+\frac{f_b}{2\pi}I_b
 -\frac{f_c}{2\pi}I_c\right) \log(\e)
 \label{4drenyi}
 \end{equation}
where $I_a=\int_{\Sigma} R_\Sigma$, $I_b=\int_{\Sigma} \tilde K_{ij}^a \tilde K_{ij}^a$ and $I_c=\int_{\Sigma} \gamma^{ij}\gamma^{kl}C_{ikjl}$ are three defect Weyl invariants built out of the 2d Ricci scalar $R_{\S}$, the traceless part of the extrinsic curvature $\tilde K_{ij}^a$ and the embedded Weyl tensor $C_{ikjl}$ contracted with the embedded metric $\gamma^{ij}$. All the integrals are performed over the defect with the appropriate invariant measure. Here we are interested in the $I_a$ invariant, a.k.a. Euler characteristics. For a spherical defect that is the only non-vanishing contribution and we have
\begin{equation}
 \int_{S^2} R_{S^2}= 8\pi
\end{equation}
Combining this with \eqref{evenpresult} we get
\begin{equation}\label{surfaceres}
 \pa_{\l} f_{a} = \frac{\p}{2} S_{q-1} C_O
\end{equation}
which is one of the main result of the paper as it predicts that the $a$-type Weyl anomaly coefficient $f_a$ is not invariant on the conformal manifold and its derivative is controlled by the one-point function of the exactly marginal operator $\mathcal{O}$. 

Looking at specific examples, the most studied one is surely the Gukov Witten defect in $\mathcal{N}=4$ SYM theory \cite{Gukov:2006jk}. In that case, there is strong evidence that the $a$-anomaly coefficient does not depend on the marginal coupling \cite{Drukker:2008wr,Jensen:2018rxu} \footnote{The statement that this coefficient vanishes, made in \cite{Drukker:2008wr}, was corrected in \cite{Jensen:2018rxu} finding agreement with the expression of the entanglement entropy.}. Using equation \eqref{evenpresult} we can actually prove this statement and make a more general one. We will show that for every surface defect in four dimensions preserving two supercharges of opposite chirality $Q_+$ and $\tilde{Q}_{\dot{-}}$, the r.h.s. of \eqref{surfaceres} vanishes. The least supersymmetric surface in this class is a $\mathcal{N}=(2,0)$ surface with superconformal algebra $su(1,1|1)\oplus sl(2)\oplus u(1)$ (see \cite{Bianchi:2019sxz} for a recent discussion). Without loss of generality we can restrict to a $\mathcal{N}=1$ bulk four-dimensional theory. The result will obviously apply to higher supersymmetric cases. The conformal manifold of a supersymmetric theory is always complex valued and the Lagrangian can be split in a chiral part belonging to a chiral multiplet and an anti-chiral part belonging to the conjugate one \cite{Green:2010da}. This supermultiplet contains a scalar operator $\mathcal{E}$ of dimension $3$ and $r$-charge $2$, a fermionic operator $\Lambda_{\alpha}$ of dimension $7/2$ and $r$-charge $1$ and the exactly marginal operator $\mathcal{O}$. It is immediate to write down the supersymmetry transformation of the fermionic operator
\begin{align}
 \d \Lambda_{\a}=\xi_{\a} \mathcal{O}+ \s^{\mu}_{\a\dot \a} \pa_\mu \mathcal{E} \tilde \xi^{\dot \a}
\end{align}
where $\xi_{\a} (\tilde \xi^{\dot \a})$ is the Grassmann odd parameter associated to the supercharge $Q_{\a}(\tilde Q_{\dot \a})$. Clearly, if the defect preserves $Q_+$ the condition $\braket{\d \Lambda_+}_{\S}=0$ readily gives $\braket{\mathcal{O}}_{\S}=0$. At the same time, if the defect preserves $U(1)$ R-symmetry $\braket{\mathcal{E}}_{\S}=0$ preventing the marginal operator from acquiring a non-vanishing one-point function anyway. This argument does not apply to superconformal defects preserving a combination of $Q$s and $\tilde{Q}$s (notice also that this is the only way for a surface defect to preserve a single supercharge). This breaks $U(1)$ R-symmetry as well as orthogonal rotations, although it may preserve a combination of them. In this case there is no symmetry setting to zero the one-point function of $\mathcal{O}$ and one should legitimately expect that $f_{a}$ depends on the coupling. Gukov Witten defects, despite breaking orthogonal rotations in general, do not belong to this class as they preserve 4 $Q$s and 4 $\tilde Q$s separately. Therefore, we find that the r.h.s. of \eqref{surfaceres} consistently vanishes for Gukov Witten defects.  Nevertheless, there is still a large class of defects for which this is not the case, including all non-supersymmetric surfaces (also when inserted in a bulk supersymmetric theory where we have examples of conformal manifolds).

\subsection{R\'enyi entropies}
In \cite{Bianchi:2015liz}, it was pointed out that twist operators in dimension higher than two can be treated as a conformal defect. This has led to several important developments \cite{Balakrishnan:2016ttg,Bianchi:2016xvf}, especially in connection with the proof of the Quantum Null Energy Condition \cite{Balakrishnan:2017bjg,Balakrishnan:2019gxl}. The R\'enyi entropy $S_n$ is related to the twist operator expectation value by \cite{Bianchi:2015liz}
\begin{equation}
 \braket{\t_n}=e^{(1-n)S_n}
\end{equation}
and its universal (regulator independent) part is given by \cite{Liu:2012eea}
\begin{equation}
 S^{\text{univ}}_n= \left\{ \begin{array}{ll} (-1)^{\frac{d}{2}} s_d \log \e & \text{even } d\\ 
                                      (-1)^{\frac{d-1}{2}} s_d & \text{odd } d
                                     \end{array}\right.
\end{equation}
This structure is clearly in agreement with our findings and we can relate the derivative of $s_d$ to the one-point function of $\mathcal{O}$
\begin{equation}
 (1-n) \pa_{\l} s_d= \left\{ \begin{array}{ll} S_{d-1} 2\, C_O  & \text{even } d\\ 
                                       S_{d-1} \p\, C_O & \text{odd } d
                                     \end{array}\right.
\end{equation}

For $d=4$, we can do something more. In that case the R\'enyi entropy is a surface defect and its universal part is described by \eqref{4drenyi} with the only difference that all the anomaly coefficients also depends on the replica number $n$. For this example, however, one additional relation between the coefficients is available \cite{Lewkowycz:2014jia}
\begin{equation}
 f_c(n)=\frac{n}{n-1} (a- f_a(n)- (n-1) \pa_n f_a (n))
\end{equation}
where $a$ is the $a$-anomaly coefficient of the bulk theory. Taking a derivative with respect to $\l$ and using the result and the conventions of \cite{Bianchi:2015liz}, relating $f_{c}(n)$ to the stress tensor one-point function we get
\begin{equation}
  \pa_\l h(n,\l)=\frac{2n}{3} \left(\frac{a}{\pi}- \pi (1-(n-1)\pa_n) C_O(n,\l)\right)
\end{equation}
which provides an intriguing relation between the stress-tensor one-point function and the marginal operator one-point function.

\section{Discussion}
In this letter we considered the effect of an exactly marginal deformation on a defect field theory. Thanks to the presence of a non-vanishing one-point function, the first order derivative of the defect expectation value with respect to the coupling is non-vanishing and it is determined by a single piece of defect CFT data. This implies a non-trivial coupling dependence for the $a$-type anomaly coefficient and for the proposed defect $C$-function. For the case of the four-dimensional Rényi entropy we also derived an equation relating the coupling derivative of the stress tensor to the one-point function of the stress tensor operator.

It would be interesting to analyze further examples, especially in the context of boundaries. It would also be important to better understand $b$-type anomalies for higher even dimensional defects and how they could be mapped to defect conformal data. Finally, one would like to find bounds on the allowed values of the $a$- and $b$-type defect anomaly coefficients (or their ratios) with and without supersymmetry.

\begin{acknowledgments}
It is a pleasure to thank Nadav Drukker, Madalena Lemos, Marco Meineri and Michelangelo Preti for very useful discussions. We would also like to thank Christopher Herzog and Itamar Shamir for important comments on the manuscript. This work is supported by the European Union’s Horizon 2020 research and innovation programme under the Marie Sklodowska-Curie grant agreement No 749909.
\end{acknowledgments}

\bibliography{biblio}

\end{document}